\documentclass[conference]{IEEEtran}
\IEEEoverridecommandlockouts
\usepackage{cite}
\usepackage{amsmath,amssymb,amsfonts}
\usepackage{algorithmic}
\usepackage{graphicx}
\usepackage{textcomp}
\usepackage{xcolor, soul}
\usepackage{siunitx}
\usepackage{mathtools}

\usepackage{threeparttable}
\usepackage{booktabs}

\usepackage{gensymb}

\usepackage[nolist]{acronym}
\usepackage{float}
\usepackage{multirow}
\usepackage{soul}
\usepackage{comment}
\usepackage{makecell} 

\usepackage[hyphens]{url}

\usepackage{orcidlink}

\usepackage{xcolor}

\usepackage[acronym]{glossaries}

\usepackage{threeparttable}
\usepackage{booktabs}
\usepackage{array}

\usepackage{amssymb}
\usepackage{pifont}
\newcommand{\todo}[1]{\textcolor{red}{#1}}                          
\newcommand{\RNum}[1]{\uppercase\expandafter{\romannumeral #1\relax}}  

\usepackage{array}
\newcolumntype{P}[1]{>{\centering\arraybackslash}p{#1}}
\newcolumntype{M}[1]{>{\centering\arraybackslash}m{#1}}

\def\BibTeX{{\rm B\kern-.05em{\sc i\kern-.025em b}\kern-.08em
    T\kern-.1667em\lower.7ex\hbox{E}\kern-.125emX}}

\makeatletter
\def\ps@IEEEtitlepagestyle{%
  \def\@oddhead{}%
  \def\@evenhead{}%
  \def\@oddfoot{%
    \hfil
    \parbox{0.95\textwidth}{%
      \centering\scriptsize
      © 2026 IEEE. Personal use of this material is permitted. Permission from IEEE must be obtained for all other uses, in any current or future media, including reprinting/republishing this material for advertising or promotional purposes, creating new collective works, for resale or redistribution to servers or lists, or reuse of any copyrighted component of this work in other works.
    }%
    \hfil
  }%
  \def\@evenfoot{}%
}
\makeatother

\begin{document}
\begin{acronym}
    \acro{us}[US]{Ultrasound}
    \acro{vns}[VNS]{Vagus Nerve Stimulation}
    \acro{avns}[aVNS]{auricular Vagus Nerve Stimulation}
    \acro{tes}[tES]{Transcutaneous electrical stimulation}
    \acro{prf}[PRF]{Pulse-Repetition Frequency}
    \acro{hr}[HR]{Heart Rate}
    \acro{hrv}[HRV]{Heart Rate Variability}
    \acro{ble}[BLE]{Bluetooth Low Energy}
    \acro{dsp}[DSP]{Digital Signal Processing}
    \acro{dd}[DD]{Drowsiness Detection}
    \acro{nn}[NN]{Neural Network}
    \acro{eeg}[EEG]{Electroencephalography}
    \acro{emg}[EMG]{Electromyography}
    \acro{ecg}[ECG]{Electrocardiogram}
    \acro{ffc}[FFC]{Flexible Flat Cable}
    \acro{ulp}[ULP]{Ultra Low Power}
    \acro{imu}[IMU]{Inertial Measurement Unit}
    \acro{fr}[FR]{Frame Rate}
    \acro{fps}[FPS]{Frames Per Second}
    \acro{pmic}[PMIC]{Power Management Integrated Circuit}
    \acro{wulpus}[WULPUS]{Wearable Ultra Low-Power Ultrasound}
    \acro{afe}[AFE]{Analog Front-End}
    \acro{rf}[RF]{Radio Frequency}
    \acro{ml}[ML]{Machine Learning}
    \acro{fpga}[FPGA]{Field-Programmable Gate Array}
    \acro{lvds}[LVDS]{Low-Voltage Differential Signalling}
    \acro{spi}[SPI]{Serial Peripheral Interface}
    \acro{mcu}[MCU]{Microcontroller Unit}
    \acro{adc}[ADC]{Analog to Digital Converter}
    \acro{hdl}[HDL]{Hardware Description Language}
    \acro{dma}[DMA]{Direct Memory Access}
    \acro{hmi}[HMI]{Human-Machine Interface}
    \acro{hd}[HD]{Hardware Design}
    \acro{ad}[AD]{Algorithm Development}
    \acro{das}[DAS]{Delay-and-Sum}
    \acro{tgc}[TGC]{Time-Gain Compensation}
    \acro{fp}[FP]{Floating Point}
    \acro{bw}[BW]{Bandwidth}
    \acro{fft}[FFT]{Fast Fourier Transform}
    \acro{ssvep}[SSVEP]{Steady State Visually Evoked Potential}
    \acro{bci}[BCI]{Brain-Computer Interface}
    \acro{pulp}[PULP]{Parallel Ultra Low Power}
    \acro{soc}[SoC]{System on Chip}
    \acro{ppg}[PPG]{Photoplethysmogram}
    \acro{loso cv}[LOSO CV]{Leave-One-Subject-Out Cross-Validation}
    \acro{cv}[CV]{Cross Validation}
    \acro{soa}[SoA]{State-of-the-Art}
\end{acronym}

\DeclareSIUnit{\rayl}{Rayl}

\newcommand{\review}[1]{\textcolor{black}{#1}}
\newcommand{\tochange}[1]{\textcolor{black}{#1}}

\title{Towards an Earpiece for Wearable and Dry Ultrasonic Vagus Nerve Stimulation
\thanks{The authors acknowledge support from the ETH Research Grant \mbox{ETH-C-01-21-2} (Project ListenToLight).}
}


\author{
\IEEEauthorblockN{
Sebastian Frey\IEEEauthorrefmark{2}\IEEEauthorrefmark{1}\orcidlink{0009-0000-6948-4363},
Farouk Ghezraoui\IEEEauthorrefmark{2},
Marc Nölke\IEEEauthorrefmark{4},
Marco Giordano\IEEEauthorrefmark{2},\\
Luca Benini\IEEEauthorrefmark{2}\IEEEauthorrefmark{3}\orcidlink{0000-0001-8068-3806},
Christoph Leitner\IEEEauthorrefmark{2}\IEEEauthorrefmark{1}\orcidlink{0000-0002-7058-7236}
\thanks{$^*$Corresponding authors: S. Frey (sefrey@iis.ee.ethz.ch), C. Leitner (christoph.leitner@iis.ee.ethz.ch).}
}
\IEEEauthorblockA{
\IEEEauthorrefmark{2}Department of Information Technology and Electrical Engineering, ETH Zurich, Switzerland\\
\IEEEauthorrefmark{4}PZN, Krakow, Poland\\
\IEEEauthorrefmark{3}Department of Electrical, Electronic, and Information Engineering, University of Bologna, Italy
}
}

\maketitle

\begin{abstract}
Ultrasound (US)-based auricular vagus nerve stimulation (aVNS) is emerging as a versatile, noninvasive modality for neuromodulation — spanning clinical therapies (e.g., pain management, depression) and general well-being applications. However, most existing US-based aVNS systems remain bulky and require wet coupling media (e.g., ultrasound gel) to achieve both spatially localized stimulation and efficient acoustic transmission. We present a wearable \review{US stimulation system combining a personalized, dry-coupled earpiece with a battery-powered, wireless US probe. The earpiece integrates a \qty{2}{\mega\hertz} piezoelectric transducer fully encapsulated in Humimic material designed for anatomical fit and stable coupling without externally applied gel. Hydrophone measurements of the fully encapsulated earpiece in water give a peak rarefactional pressure of \qty{0.24}{\mega\pascal} at \qty{15}{\milli\meter} for a \qty{15}{\volt} unipolar excitation.} We further demonstrate in vivo feasibility through fully dry, in-ear \review{\qty{2}{\mega\hertz}} pulse-echo acquisitions under realistic wearing conditions, confirming effective US transmission without the need for coupling gel. \review{These results demonstrate the integration of a personalized dry acoustic interface with battery-powered stimulation electronics, providing a wearable platform for investigating US-based aVNS.} Future studies will focus on functional outcomes, including heart rate variability and pupillary responses during stimulation.

\end{abstract}





\section{Introduction}

\ac{vns} is an established neuromodulation technique for diverse therapeutic applications, including epilepsy \cite{gouveia2024neurostimulation}, depression \cite{conway2025vagus}, and inflammatory conditions \cite{drewes2021short}. Traditional \ac{vns} is invasive, requires surgical implantation of electrodes and hence poses risks and limits patient accessibility \cite{korupolu2024neurorehabilitation}. To address these drawbacks, noninvasive alternatives have been explored \cite{li2022non}. For instance, \ac{tes} at the cymba concha targets the auricular branch of the vagus nerve, enabling \ac{avns} without surgery. It has shown promise in the treatment of diseases \cite{j:wang21} and in applications relating to lifestyle and well-being \cite{j:Ferstl22}. \ac{tes}, however, often induces undesirable sensations (e.g., skin irritation, pain) caused by current injection and suffers from variable electrode–skin impedances \cite{gerges2024clinical}. \ac{us}, on the other hand, is a further viable, non-invasive, \review{non-ionizing} alternative to traditional \ac{vns} \review{that penetrates skin and cartilage and can be spatially confined at depth.} This acoustic modality stimulates neural tissue via mechanical perturbation rather than electrical current \cite{downs2018}. However, to ensure effective transmission of sound waves from the transducer into human tissue, \ac{us} requires an air-free interface to prevent reflections caused by impedance mismatches. For example, Hacker et al. \cite{hacker_development_2024} developed a portable headset that integrates electronics and a transducer array for \ac{us}-based \ac{avns} in the concha of the ear. While the design incorporates the driving circuitry, its rigid form factor prevents anatomical conformity. Moreover, it requires a headset and coupling gel to maintain contact pressure and guarantee acoustic coupling. This limits long-term wearability and user comfort. Van Damme et al.~\cite{van_damme_feasibility_2024} demonstrated a flexible, high-frequency \ac{us} cuff for cervical nerve stimulation, showcasing \ac{us}-based neuromodulation. However, their design is neither anatomically optimized for auricular stimulation nor independent of external support structures, restricting its suitability for prolonged wear.
These studies highlight a critical gap: the absence of a compact, conformal, and fully dry \ac{us} delivery system optimized for auricular stimulation.

To address this gap, \review{we propose a personalized, dry-coupled earpiece integrated with battery-powered stimulation electronics and wireless control. The earpiece incorporates a miniature \qty{2}{\mega\hertz} piezoelectric transducer fully encapsulated in soft Humimic material. A vacuum-assisted molding process reproduces the individual's ear geometry while minimizing trapped air within the encapsulation. The resulting interface is designed to provide conformal transducer--skin contact to ensure comfort and enable stable acoustic coupling suitable for extended wear without externally applied coupling gel or a supporting headset. Embedded waveform generation provides a \qty{2}{\mega\hertz} carrier gated at a \qty{40}{\hertz} burst repetition rate, enabling wearable stimulation without an external waveform generator or wired power supply.}

Our key contributions are:  
\begin{enumerate}  
  \item Design and fabrication of a soft, custom-molded earpiece embedding a miniature \review{\qty{2}{\mega\hertz}} transducer with dry acoustic coupling.  
  \item \review{Integration of the dry-coupled earpiece with battery-powered stimulation electronics and wireless control, delivering a \qty{2}{\mega\hertz} carrier gated at a \qty{40}{\hertz} burst repetition rate.}
 \item \review{Hydrophone-based characterization of the fully encapsulated earpiece, demonstrating a maximum measured peak-to-peak acoustic pressure of \qty{0.47}{\mega\pascal} in water.}
  \item \emph{In vivo} demonstration of fully dry, in-ear pulse-echo \ac{us} performance under realistic wearing conditions.  
\end{enumerate}

\section{Materials and Methods}
\label{sec:methods}

\begin{figure}[t]
    \centering
    \includegraphics[trim={18pt 0 0 0}, clip, width=\linewidth]{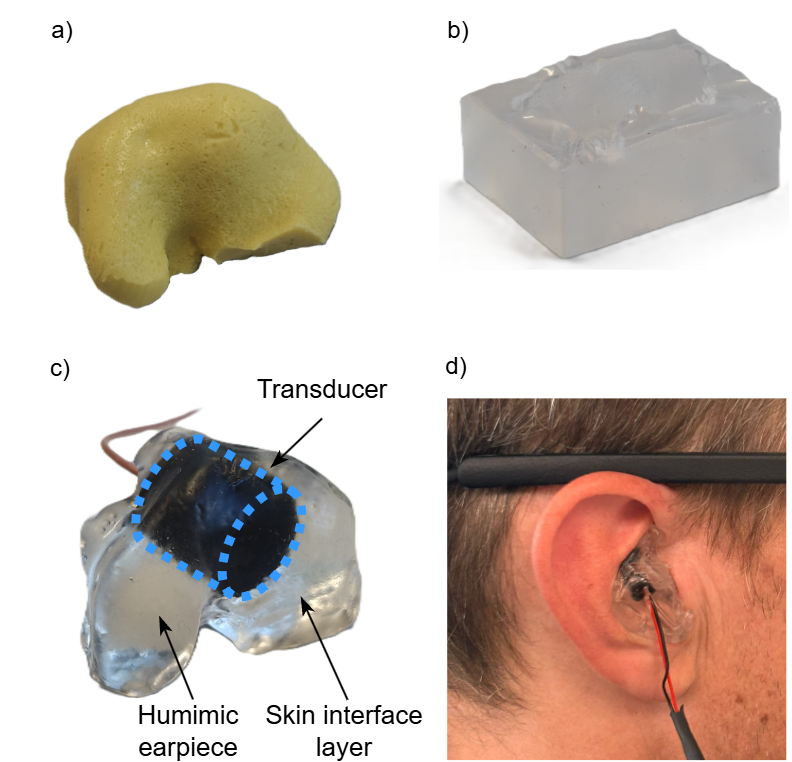}
    \caption{\review{Overview of the earpiece design process. a) Personalized ear impression. b) Flexible Ecoflex mold used to facilitate earpiece demolding. c) Humimic earpiece with the integrated transducer and coupling layer. d) Subject wearing the earpiece with the transducer directed towards the cymba concha to target the auricular branch of the vagus nerve.}}
    \label{fig:earpiece_design_steps}
\end{figure}

\subsection{Earpiece Design and Fabrication}
\review{A custom wearable \ac{us} earpiece was developed for noninvasive \ac{avns}. The fabrication process started from a personalized ear impression (Figure~\ref{fig:earpiece_design_steps}a), which was used to produce a flexible Ecoflex mold (Figure~\ref{fig:earpiece_design_steps}b). The resulting Humimic earpiece fully encapsulates a \qty{2}{\mega\hertz} piezoelectric transducer (Piezo Hannas, China), with a continuous layer of material covering its emitting face to form the skin-contact acoustic interface (Figure~\ref{fig:earpiece_design_steps}c). The personalized geometry is designed to conform to the auricle and maintain acoustic contact without externally applied coupling gel, with the transducer oriented toward the cymba concha when worn (Figure~\ref{fig:earpiece_design_steps}d).}

\begin{figure*}[bth]
    \centering
    \includegraphics[width=\linewidth]{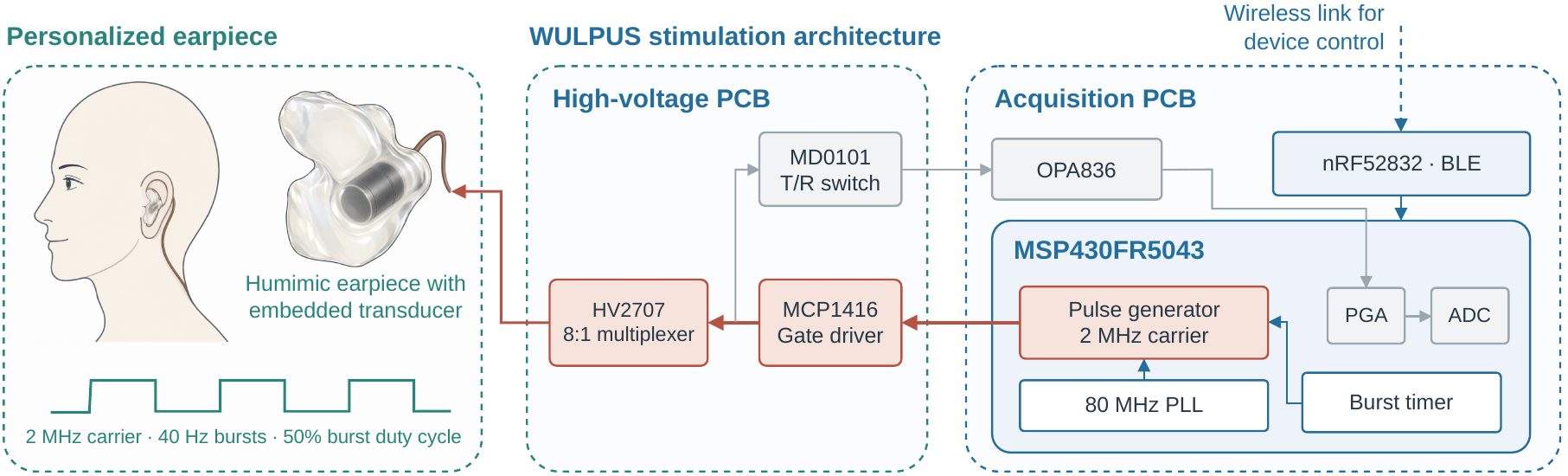}
    \caption{\review{Overview of the wearable \ac{us} stimulation system. The battery-powered WULPUS platform generates a \qty{2}{\mega\hertz} carrier using the MSP430FR5043 microcontroller and drives the transducer embedded within the personalized Humimic earpiece through its high-voltage circuitry. Carrier gating produces a \qty{40}{\hertz} burst repetition rate at a \qty{50}{\percent} burst duty cycle.}}
    \label{fig:earpiece_concept}
\end{figure*}
\review{A personalized ear impression was first obtained and used to produce a negative mold from Ecoflex silicone. This flexible mold reproduced the individual's auricular geometry and facilitated removal of the finished earpiece. Before casting, the transducer was positioned within the mold with its emitting face oriented toward the region corresponding to the cymba concha. Space was retained between the transducer face and the mold surface to form a continuous Humimic coupling layer over the emitting face.}

Humimic gel (Humimic Medical, USA) was selected to fabricate the earpiece body and the acoustic coupling layer covering the transducer. Its softness and thermal moldability enable the fabrication of a conformal interface with the transducer embedded directly within the material. The material properties are summarized in Table~ \ref{table:earpiece_properties} \cite{w:humimic2025}.

\begin{table}[h!]
\centering
\caption{Material properties of the earpiece.}
\label{table:earpiece_properties}
\resizebox{\columnwidth}{!}{%
\begin{threeparttable}
\begin{tabular}{p{1.6in}>{\arraybackslash}p{1.4in}}
\toprule[0.20em]
\textbf{Earpiece Material Property} & \textbf{Value} \\
\midrule
Young’s Modulus & 0.19 MPa\\
Speed of Sound & 1449.30 $\pm$ 3.37 m/s\\
Acoustic Impedance & 1.37 $\pm$ 0.002 MRayl\\
Acoustic Attenuation, $\alpha$ & 0.223 $\pm$ 0.002 dB/(MHz cm)\\
Acoustic Attenuation, $\nu$ & 1.7 $\pm$ 0.04 \\
\bottomrule[0.20em]
\end{tabular}
\end{threeparttable}
}
\end{table}

\review{With the transducer positioned inside the Ecoflex mold, pieces of Humimic material were placed above the mold cavity. The assembly was placed in a vacuum chamber, evacuated, and heated to \qty{120}{\degreeCelsius} for \qty{1}{\hour}, allowing the Humimic to melt, fill the cavity, and surround the transducer. Vacuum-assisted processing was used to minimize trapped air within the encapsulation, particularly along the acoustic path in front of the transducer. The assembly was subsequently allowed to cool until the Humimic solidified. The earpiece was then demolded, and its surface was locally smoothed using hot air. The resulting structure fully enclosed the transducer, including its emitting face, while allowing the electrical cable to exit the earpiece.}

\review{\subsection{Wearable Stimulation System}}
\review{As shown in Figure~\ref{fig:earpiece_concept}, the wearable stimulation electronics were implemented using the WULPUS A-mode \ac{us} platform \cite{frey_wulpus_2022}, which provides battery operation and wireless control and connects by cable to the encapsulated transducer. Its acquisition board integrates an nRF52832 microcontroller (Nordic Semiconductor) for wireless communication, synchronization, and command reception, and an MSP430FR5043 microcontroller (Texas Instruments) for local waveform generation. A stacked high-voltage board incorporates a MOSFET gate driver (MCP1416, Microchip) and multiplexer (HV2707, Microchip) to deliver unipolar excitation pulses of \qty{15}{\volt} to the transducer. The MSP430's integrated ultrasonic sensing module generates the \qty{2}{\mega\hertz} carrier using a programmable pulse generator clocked at \qty{80}{\mega\hertz} by a phase-locked loop (PLL), referenced to an external \qty{8}{\mega\hertz} ceramic resonator. A hardware timer interrupt alternates the carrier between \qty{12.5}{\milli\second} ON and OFF intervals, producing a \qty{40}{\hertz} burst repetition rate and a \qty{50}{\percent} burst duty cycle. The burst parameters were selected with reference to previous \ac{us}-based \ac{avns} work \cite{hacker_development_2024}. Between bursts, the output is actively driven low, while the ultrasonic sensing module and PLL remain enabled to avoid repeated startup delays.}

\subsection{Acoustic \review{Output Characterization}}
Hydrophone measurements were performed \review{in a tank filled with deionized, degassed water to characterize the acoustic output of the complete Humimic-encapsulated earpiece. The transducer was driven by the wearable stimulation electronics using unipolar pulses from \qty{0}{\volt} to \qty{15}{\volt} at a carrier frequency of \qty{2}{\mega\hertz}. A calibrated needle hydrophone (HP33825, Precision Acoustics, UK), with a sensitivity of \qty{45}{\milli\volt\per\mega\pascal} at \qty{2}{\mega\hertz}, was positioned along the transducer's nominal acoustic axis. Measurements were recorded at distances of \qtylist{5;10;15}{\milli\meter} from the outer Humimic surface. Peak-to-peak acoustic pressure was calculated by dividing the measured peak-to-peak hydrophone voltage by the hydrophone sensitivity. Assuming a symmetric sinusoidal progressive wave, the peak rarefactional pressure was estimated as $p_r=p_{\mathrm{pp}}/2$, and the pulse-average intensity as $I_{\mathrm{PA}}=p_r^2/(2Z)$, where $Z$ is the acoustic impedance of water (\qty{1.5e6}{\kilogram\per\meter\squared\per\second}). Temporal-average intensity was calculated as $I_{\mathrm{TA}}=D I_{\mathrm{PA}}$, with burst duty cycle $D=0.5$.}


\subsection{\ac{us} In-Ear Pulse-Echo}

The earpiece was placed in the participant's concha (Figure \ref{fig:earpiece_design_steps}d). Written informed consent was obtained. The transducer, fully enclosed within the earpiece, is located in its superior section and oriented toward the cymba concha. This region is densely innervated by the auricular branch of the vagus nerve, which runs through the cartilage and skin at a depth of approximately 1 – \qty{1.5}{\milli\meter} \cite{kaniusas2019current}. The Humimic material provides structural support and serves as a soft, conformable acoustic coupling layer, ensuring efficient \ac{us} transmission by minimizing air gaps at the interface.

\subsubsection{Measurement Protocol}
\ac{us} signal envelopes were acquired using \textit{ModulUS} \cite{c:modulusIUS25}, an in-house platform for wearable \ac{us} research. In three consecutive sessions, we collected \ac{us} pulse-echo data at an ultrasound frequency of \qty{11}{\mega\hertz} and a \ac{prf} of \qty{25}{Hz} for \qty{60}{\second}, resulting in 4500 envelope scanlines. 

\subsubsection{Data Processing}
Data were processed in Python. First, the raw \ac{us} envelope signals were time-gain compensated using the implementation provided in \textit{dasIT} \cite{leitner2025dasIT}. For visualization, the DC bias was removed, and the data were smoothed using a Gaussian kernel (FWHM = 150). Peaks were detected using a prominence threshold of 0.05 and a height threshold of 0.02.

\section{Results}

\subsection{Acoustic Output Characterization}
\review{When driven with unipolar pulses from \qty{0}{\volt} to \qty{15}{\volt} at a carrier frequency of \qty{2}{\mega\hertz}, the fully encapsulated transducer produced hydrophone voltages of \qtylist{15.1;16.5;21.3}{\milli\volt} peak-to-peak at distances of \qtylist{5;10;15}{\milli\meter} from the outer Humimic surface, respectively. These correspond to peak-to-peak acoustic pressures of \qtylist{0.34;0.37;0.47}{\mega\pascal}. The maximum measured pressure occurred at \qty{15}{\milli\meter}. Assuming a symmetric sinusoidal progressive wave and an acoustic impedance of water of \qty{1.5}{\mega\rayl}, the estimated peak rarefactional pressure at this location was \qty{0.24}{\mega\pascal}, with a pulse-average intensity of \qty{1.87}{\watt\per\centi\meter\squared}. At a \qty{50}{\percent} burst duty cycle, the corresponding temporal-average intensity was \qty{0.93}{\watt\per\centi\meter\squared}. These values characterize the acoustic output of the complete wearable assembly in water, including transmission through the Humimic encapsulation, without tissue derating.}

\begin{figure}[bth]
    \centering
    \includegraphics[width=\linewidth]{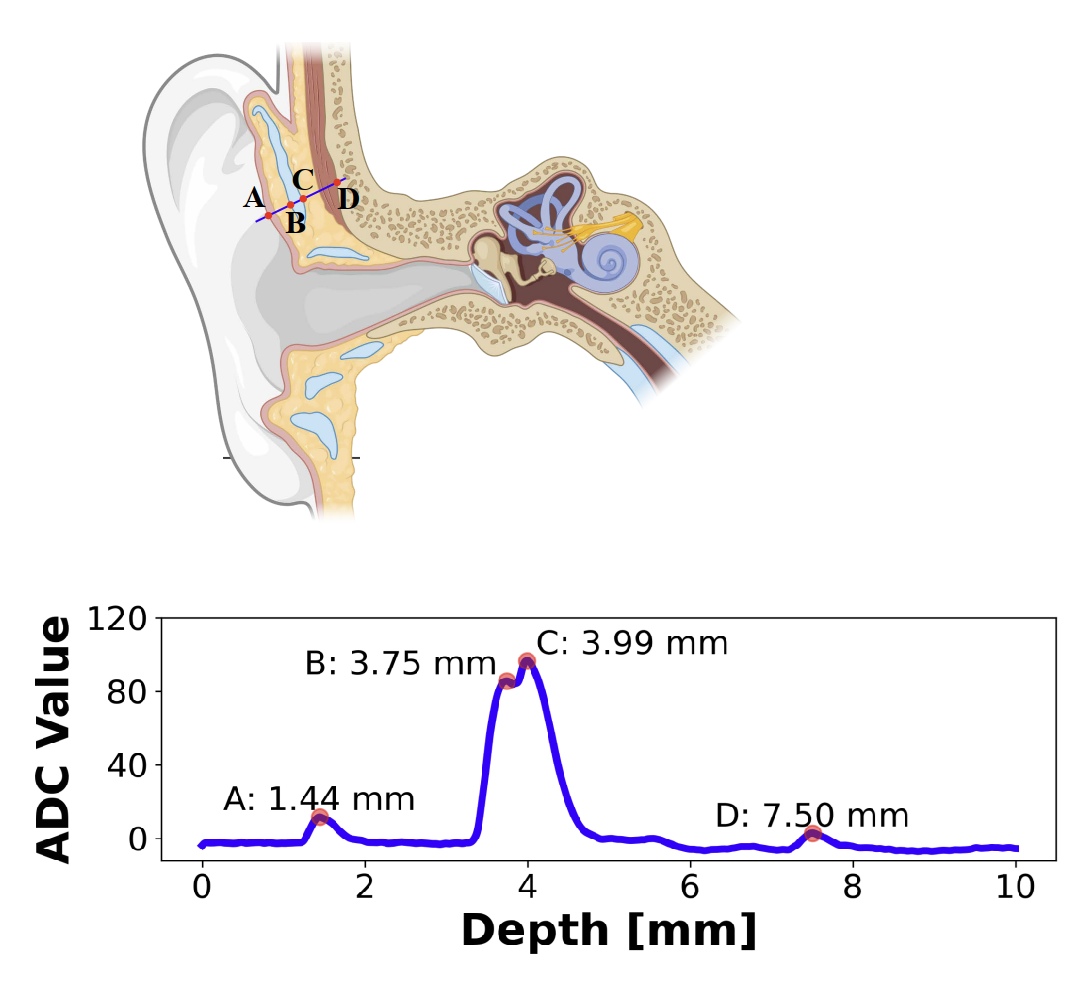}
    \caption{Pulse-echo envelope signals acquired with our custom earpiece. Reflective targets at the skin surface (A), cartilage boundaries (B–C), and cortical bone (D) are clearly visible in the \ac{us} envelopes, demonstrating the system's capability to operate in a dry configuration.}
    \label{fig:pulseecho}
\end{figure}

\subsection{\ac{us} In-Ear Pulse-Echo}
Figure~\ref{fig:pulseecho} shows a pulse-echo \ac{us} recording acquired with our custom earpiece. Reflective targets at the skin surface (A), cartilage boundaries (B–C), and cortical bone (D) are clearly visible in the \ac{us} envelopes. Signal amplitudes degrade with depth, most likely due to the absence of a variable gain amplifier in the system.



\section{Conclusion}
\review{This study presented a personalized dry acoustic interface integrated with battery-powered, wirelessly controlled electronics for wearable \ac{us}-based \ac{avns}. Vacuum-assisted molding enabled full encapsulation of a \qty{2}{\mega\hertz} piezoelectric transducer within a conformal Humimic earmold. Local waveform generation and transducer driving circuitry provided a \qty{2}{\mega\hertz} carrier gated at \qty{40}{\hertz}, enabling operation without an external signal generator or wired power supply. Hydrophone measurements demonstrated a maximum measured peak-to-peak acoustic pressure of \qty{0.47}{\mega\pascal} in water, while in-ear pulse-echo acquisitions demonstrated acoustic transmission without externally applied coupling gel. Together, these results establish an integrated platform for investigating dry, wearable \ac{us}-based \ac{avns}.}
Future work will include in vivo physiological testing, focusing on heart rate variability and pupillary responses to evaluate functional effects.

\section*{Acknowledgment}
We thank H. Gisler and T. Quanbrough (ETH Z{\"u}rich) for technical support. 

\bibliographystyle{IEEEtran}
\bibliography{bibliography.bib}

\end{document}